\documentclass[fleqn,usenatbib]{mnras}

\usepackage[T1]{fontenc}

\DeclareRobustCommand{\VAN}[3]{#2}
\let\VANthebibliography\thebibliography
\def\thebibliography{\DeclareRobustCommand{\VAN}[3]{##3}\VANthebibliography}

\usepackage{graphicx}	
\usepackage{amsmath}	
\usepackage{amssymb}	

\usepackage{newtxtext,newtxmath} 

\newcommand{\be}{\begin{equation}}
\newcommand{\ee}{\end{equation}}
\newcommand{\bea}{\begin{eqnarray}}
\newcommand{\eea}{\end{eqnarray}}

\title[From Hubble to Snap Parameters]{From Hubble to Snap Parameters:
A Gaussian Process Reconstruction}

\author[J. F. Jesus et al.]{
J. F. Jesus,$^{1,2}$\thanks{E-mail: jf.jesus@unesp.br (JFJ)}
D. Benndorf,$^{2}$\thanks{E-mail: douglas.benndorf@unesp.br (DB)}
A. A. Escobal$^{2}$\thanks{E-mail: anderson.escobal@unesp.br (AAE)}
and S. H. Pereira$^{2}$\thanks{E-mail: s.pereira@unesp.br (SHP)}
\\
$^1$Universidade Estadual Paulista (UNESP), Instituto de Ciências e Engenharia, R. Geraldo Alckmin, 519, 18409-010, Itapeva, SP, Brazil,
\\$^2$Universidade Estadual Paulista (UNESP), Faculdade de Engenharia e Ciências de Guaratinguet\'a,\\ Departamento de F\'isica - Av. Dr. Ariberto Pereira da Cunha 333, 12516-410, Guaratinguet\'a, SP, Brazil
}

\date{Accepted XXX. Received YYY; in original form ZZZ}

\pubyear{2023}

\begin{document}
\label{firstpage}
\pagerange{\pageref{firstpage}--\pageref{lastpage}}
\maketitle

\begin{abstract}
By using recent $H(z)$ and SNe Ia data, we reconstruct the evolution of kinematic parameters $H(z)$, $q(z)$, jerk and snap, using a model-independent, non-parametric method, namely, the Gaussian Processes. Throughout the present analysis, we have allowed for a spatial curvature prior, based on Planck 18 constraints. In the case of SNe Ia, we modify a python package (GaPP) in order to obtain the reconstruction of the fourth derivative of a function, thereby allowing us to obtain the snap from comoving distances. Furthermore, using a method of importance sampling, we combine $H(z)$ and SNe Ia reconstructions in order to find joint constraints for the kinematic parameters. We find for the current values of the parameters: $H_0 =67.2 \pm 6.2$ km/s/Mpc, $q_0 = -0.54^{+0.06}_{-0.05}$, $j_0=0.94^{+0.20}_{-0.18}$, $s_0=-0.62^{+0.26}_{-0.25}$ at 1$\sigma$ c.l. We find that these reconstructions are compatible with the predictions from flat $\Lambda$CDM model, at least for 2$\sigma$ confidence intervals.
\end{abstract}




\section{Introduction}


The relatively recent discovery of an accelerated expansion of the universe is confirmed by observations of Supernovae Type Ia (SNe Ia) (\cite{pantheonplus})
and differential age of distant galaxies, through Hubble parameter ($H(z)$) measurements (\cite{MaganaEtAl17}). It is well known that the standard model of cosmology, $\Lambda$CDM, fits quite well the observational data, however, alternative models have also been proposed recently to deal with some problems suffered by the standard model (\cite{Bull2016}). Just to cite some examples of such alternative models, the dark energy and dark matter effects, or even the interaction among them, sometimes are inserted into the dynamic equations as new matter components  (\cite{GongBo,marttens2020,SF, Majerotto2009,Valiviita2010,Chimento2010,Cai2010,Sun2012,Pourtsidou2013,Salvatelli2014,Li2014,Skordis2015,Jimenez2016,Valent2020}).

Another way to study the cosmic evolution is through the so called cosmographic or kinematic models (\cite{kine1,kine2,kine3,kine4,kine5,kine6,kine7}), where one is not worried about the Universe composition, but instead, how it evolves. In such models we seek for direct measures of expansion through its kinematic parameters (such as the Hubble parameter $H_0$, deceleration parameter $q$, jerk $j$ and snap $s$ parameters, etc). The main advantage of the method is to obtain a direct measure of expansion parameters directly from the data, with fewer assumptions than in dynamic modelling.

Recently, different types of cosmographic dark energy models have been studied. Running vacuum models was studied by \cite{rezaei2021}, the study of cosmographic parameters in model independent approaches was done in \cite{mehrabi2021}, cosmographic functions up to the fourth order of derivative of the scale factor with the non-parametric method of Gaussian Processes was done by \cite{velasques2021}, among others.

One way to implement cosmographic modeling is to parameterize the cosmological parameters $H_0$, $q$, $j$ and $s$ for instance, or even higher orders if desired (crackle, pop, etc) (\cite{lobo2020}). Another way is to obtain the cosmological parameters via reconstruction by Gaussian processes. 
 \cite{Bilicki:2012} reconstructed $q(z)$, the normalized Hubble parameter ($E(z)$) and its two derivatives with respect to redshif $z$. 
  \cite{NanLin:2019} reconstructed the normalized comoving distance $d_C(z)$, $E(z)$, $q(z)$ and the equation of state parameter $w(z)$. 
 \cite{Zhang16} reconstructed $q(z)$ and the dimensionless comoving luminosity distance $D(z)$ (and its derivatives). \cite{Haridasu18} reconstructed $q(z)$ and $E(z)$.  Mukherjee and Banerjee reconstructed $q(z)$, $H(z)$, $D_C(z)$ (with two derivatives) (\cite{Mukherjee20}) and $j(z)$ (\cite{Mukherjee21}).

The idea of Gaussian Processes (GP) is to obtain from a dataset $f(x_i)\pm\sigma_i$ a reconstruction of the function $f(x)$. This is done by assuming that the errors in the data are Gaussian. In this way, the underlying function that describes the data is reconstructed as a Gaussian process, that is, it is generated from a point-to-point Gaussian distribution.

{It was shown by } \cite{SeikelEtAl12a,SeikelEtAl12b}, \cite{YuEtAl18} and \cite{JesusEtAl20GP} that the expected value $\mu$ and the variance $\sigma^2$ of the function $f(x)$ in the GP method are given by:
\begin{align}
    \mu(x)&=\sum_{i,j=1}^Nk(x,x_i)(M^{-1})_{ij}f(x_j)\label{mux}\\
    \sigma^2(x)&=k(x,x)-\sum_{i,j=1}^Nk(x,x_i)(M^{-1})_{ij}k(x_j,x)\label{varx}
\end{align}
where $N$ is the number of data, the matrix $M_{ij}=k(x_i,x_j)+c_{ij}$, $c_{ij}$ is the covariance matrix of the data and $k(x, x')$ is the covariance function or kernel between the points $x$ and $x'$.

One of the main ingredients to obtain the GP reconstruction is the covariance function. As we assume that the data describe the same underlying $f(x)$ function, two points $x$ and $x'$ are not independent of each other, so they have a covariance. To perform the reconstruction, it is necessary to choose a covariance function, the function that will model the correlation between these points. A commonly used covariance function is the quadratic exponential (or Gaussian) function:
\begin{equation}
    k(x,x')=\sigma_f^2\exp\left[-\frac{(x-x')^2}{2l^2}\right]
\end{equation}
where $\sigma_f$ and $l$ are hyperparameters of the covariance function. $\sigma_f$ is related to the vertical variation of the data and $l$ is related to the horizontal variation. {We have chosen the quadratic exponential kernel because it is a smooth function, infinitely differentiable, which is interesting for the present work, where we need to reconstruct up to the 4th derivatives.}

In order to use the equations \eqref{mux} and \eqref{varx} to reconstruct the function $f(x)$, we need to determine the hyperparameters. They can be obtained from maximizing the logarithmic marginalized likelihood function:
\begin{equation}
\ln\mathcal{L}=-\frac{1}{2}\sum_{i,j=1}^Nf(x_i)\left(M^{-1}\right)_{ij}f(x_j)-\frac{1}{2}\ln|M|-\frac{1}{2}N\ln2\pi
\end{equation}
where $|M|$ is the determinant of $M_{ij}$.

In this work we reconstruct the cosmological parameters $H(z)$, $q(z)$, $j(z)$ and $s(z)$ via the Gaussian Processes from SNe Ia data and $H(z)$ data. First we make the separate analysis for each set of data and then its combination. 

The paper is organised as follows: In Section II we present the equations for the kinematic parameters. Section III contain the analyses and results and Section IV, the conclusions.

\section{Equations for kinematic parameters}

The cosmological kinematic parameters are given by:
\begin{align}
    H&= \frac{\dot{a}}{a}\label{Hdef}\\
    q&=-\frac{\ddot{a}}{aH^2}\label{qdef}\\
    j&= \frac{\dddot{a}}{aH^3}\label{jdef}\\
    s&= \frac{\ddddot{a}}{aH^4}\label{sdef}
\end{align}
for Hubble, deceleration, jerk and snap parameters, respectively, where $a(t)$ is the scale factor of Friedmann-Robertson-Walker metric, $\dot{a} \equiv da/dt$ and $H = \dot{a}/a$. The negative sign on the deceleration parameter $q$ is due to historical reasons.

\subsection{Kinematic Parameters from \texorpdfstring{$H(z)$}{Hz}}

The Gaussian Process (GP) will reconstruct $H(z)$ and its derivatives from $H(z)$ data and $D_M(z)$ from SNe Ia data. Let us then write the kinematic parameters in terms of $H(z)$ and its derivatives. First of all, let us realize that $\dot{a}=aH$ and $\frac{d}{dt}=-H(1+z)\frac{d}{dz}$. Thus:
\begin{align}
q=-\frac{1}{aH^2}\frac{d(aH)}{dt}=
-1+(1+z)\frac{H'(z)}{H(z)}
\label{qzdef}
\end{align}

Let us now find an expression for the jerk. As shown in \cite{BenndorfEtAl22}, the jerk can be obtained from the deceleration parameter as
\be
j(z)=(1+z)q'(z)+q(1+2q)
\ee
Thus, from \eqref{qzdef}, we have
\be
j(z)=1-2(1+z)\frac{H'(z)}{H(z)}+(1+z)^2\frac{H'(z)^2}{H(z)^2}+(1+z)^2\frac{H''(z)}{H(z)}
\label{jzdef}
\ee

Finally, as shown in \cite{BenndorfEtAl22}, the snap can be obtained as
\be
s(z)=-(1+z)j'(z)-\left[2+3q(z)\right]j(z)
\label{sjq}
\ee
So, from Eqs. \eqref{qzdef} and \eqref{jzdef}, we find
\begin{align}
s(z)=&1-3(1+z)\frac{H'}{H}+3(1+z)^2\frac{H'^2}{H^2}-(1+z)^3\frac{ H'^3}{H^3}+\nonumber\\
&-4(1+z)^3\frac{H'H''}{H^2}+(1+z)^2\frac{H''}{H}-(1+z)^3\frac{H'''}{H}
\end{align}
where the primes denote derivatives with respect to the redshift $z$.

\subsection{Kinematic Parameters from \texorpdfstring{$D_M(z)$}{DMz}}
In order to use the reconstruction made by the GPs for the SNe Ia data, let us write the kinematic parameters in terms of the dimensionless transverse comoving distance $D_M(z)$, which is given by
\be
D_M(z)=\frac{1}{\sqrt{-\Omega_k}}\sin\left(\sqrt{-\Omega_k}D_C(z)\right)
\label{DMDC}
\ee
where the line-of-sight comoving distance $D_C(z)$ relates to $E(z)\equiv\frac{H(z)}{H_0}$ as:
\be
D_C'(z)= \frac{1}{E(z)}\,,
\label{DCEz}
\ee
where a prime denote a derivative with respect to redshift $z$. Dimensionless distances $D_i$ relate to dimensionful distances $d_i$ as:
\be
D_i\equiv\frac{d_i}{d_H},
\ee
where $d_H\equiv\frac{c}{H_0}$ is Hubble distance.

The derivative with respect to $z$ of the transverse comoving distance $D_M(z)$ is
\be
\dfrac{d D_M(z)}{dz}=D_M'(z)=\cos\left(\sqrt{-\Omega_k}D_C(z)\right)D_C'(z)\,,
\label{dDMDC}
\ee
so, by combining \eqref{DMDC} and \eqref{dDMDC}, we find:
\be\label{EDM}
\left(\frac{D_M'}{D_C'}\right)^2-\Omega_kD_M^2=1\,.
\ee
Now, by using \eqref{DCEz}, we can express the relation \eqref{EDM} in terms of the dimensionless quantity $E(z)$:
\be
E^2=\frac{1+\Omega_kD_M^2}{D_M'^2}\,.
\label{E2DM}
\ee
Taking the derivative with respect to $z$ of the relation \eqref{E2DM}, we have
\be
E\frac{dE}{dz}=\frac{\Omega_kD_M\left(D_M'^2-D_MD_M''\right)-D_M''}{D_M'^3}\,,
\label{EdEdz}
\ee

By dividing \eqref{EdEdz} by \eqref{E2DM} and using \eqref{qzdef}, we find

\be
q(z)=\frac{\Omega_k(1+z) D_M(z) D_M'(z)}{1+\Omega_kD_M(z)^2}-(1+z)\frac{ D_M''(z)}{D_M'(z)}-1
\label{qzDc}
\ee

Similarly, the jerk is given by
\begin{align}
j(z)=&1+\frac{\Omega_k(1+z)^2}{1+\Omega_kD_M^2}
\left[D_M'^2-\frac{2D_MD_M'}{1+z}-3D_MD_M''\right] \nonumber \\
&+\frac{(1+z)^2}{D_M'}\left[\frac{3 D_M''^2}{D_M'}+\frac{2D_M''}{(1+z)}-D_M'''\right]
\label{jzDc}
\end{align}

As we can see from these equations, one can always separate them in a part that depends on $\Omega_k$ and one that does not. Let us separate them in order to simplify equations. So, \eqref{qzDc} can be written as:
\be
q(z)=q_f(z)+q_k(z)
\ee
where
\begin{align}
q_f(z)&\equiv-1-(1+z)\frac{ D_M''(z)}{D_M'(z)}\\
q_k(z)&\equiv\frac{\Omega_k(1+z) D_M(z) D_M'(z)}{1+\Omega_kD_M(z)^2}
\end{align}

Similarly,
\be
j(z)=j_f(z)+j_k(z)
\ee
where
\begin{align}
j_f(z)&\equiv1+\frac{(1+z)^2}{D_M'}\left[\frac{3 D_M''^2}{D_M'}+\frac{2D_M''}{(1+z)}-D_M'''\right]\\
j_k(z)&\equiv\frac{\Omega_k(1+z)^2}{1+\Omega_kD_M^2}
\left[D_M'^2-\frac{2D_MD_M'}{1+z}-3D_MD_M''\right]
\end{align}

Using \eqref{sjq}, then, we have:
\be\label{szDc}
s(z)=-(1+z)j_f'-(2+3q_f)j_f-(1+z)j_k'-(2+3q_f+3q_k)j_k-3j_fq_k
\ee

From where we can identify:
\begin{align}
s_f(z)&\equiv-(1+z)j_f'-(2+3q_f)j_f\\
s_k(z)&\equiv-(1+z)j_k'-(2+3q_f+3q_k)j_k-3j_fq_k
\end{align}

So:
\begin{align}
s_f(z) &=1+ \frac{(1+z)^3 D_M^{(4)}}{D_M'} - \frac{(1+z)^2 D_M^{(3)}}{D_M'} + \frac{15 (1+z)^3D_M''^3}{D_M'^3} + \nonumber\\
&+\frac{5 (1+z)^2 D_M''^2}{D_M'^2} + \frac{3 (1+z)D_M''}{D_M'} - \frac{10 (1+z)^3 D_M^{(3)} D_M''}{D_M'^2}
\label{sf}
\end{align}

\begin{align}\nonumber
s_k=&\frac{\Omega_k (1+z)}{\left(1+\Omega_k D_M^2\right)^2 D_M'} \left\{\Omega_k (1+z) D_M^2 D_M'^2 \left(7 (1+z) D_M''+3D_M'\right) \right.+\\\nonumber&+(1+z) D_M'^2 \left(4 (1+z) D_M''+D_M'\right)+\\\nonumber&-\Omega_k
   D_M^3 \left[18 (1+z)^2 D_M''^2+3 D_M'^2+\right.\\\nonumber&\left. +(1+z) D_M' \left(7 D_M''-6
   (1+z) D_M^{(3)}\right)\right]+\\\nonumber&-D_M \left[18 (1+z)^2 D_M''^2+\Omega_k (1+z)^2
   D_M'^4+3 D_M'^2\right.+\\  &+ (1+z)\left. D_M'\left. \left(7 D_M''-6 (1+z)
   D_M^{(3)}\right)\right]\right\}\,.
\end{align}


With these expressions, we can reconstruct the kinematic parameters evolution from $H(z)$ and SNe Ia data.

\section{Analyses and Results}
The observational data sample used in this work is formed by { 1701 light curves of $1550$ distinct SNe Ia from the Pantheon+ sample (\cite{pantheonplus})  }and a compilation of $32$ Hubble parameter data, $H(z)$ (\cite{MorescoEtAl22}), obtained by estimating the differential ages of galaxies, called cosmic chronometers.

Concerning the Pantheon sample, we have followed the same idea as in \cite{JesusEtAl20GP,JesusEtAl22GP}. As explained by \cite{JesusEtAl20GP}, as the Pantheon sample consists of apparent magnitude SNe Ia data, this is not suitable for GP reconstruction, as magnitudes diverge for $z\rightarrow0$ and GP fails to reconstruct rapidly varying functions as explained by \cite{SeikelEtAl12a}. Thus, we made a error propagation from magnitudes to comoving distances, as explained in \cite{JesusEtAl20GP}.

The 32 $H(z)$ Cosmic Chronometers (CCs) data is a sample compiled by \cite{MorescoEtAl22}, where they have, for the first time, estimated systematic errors for these data. In order to do that, they have made simulations and have considered effects as metallicity, rejuvenation effect, star formation history, initial mass function, choice of stellar library etc.\footnote{The method to obtain the full covariance matrix, together with jupyter notebooks as examples is furnished by M. Moresco at \url{https://gitlab.com/mmoresco/CCcovariance}.} {When such systematic errors are taken into account, the $H(z)$ data errors are increased, which is in line with the claims by \cite{Kjerrgren:2021zuo} }

We have used the publicly available software GaPP by \cite{SeikelEtAl12a} in order to make the reconstructions in the present work. It is important to mention, however, that the original software were able only to reconstruct function derivatives up to third order, while to obtain the snap reconstruction from SNe Ia, we need the fourth derivative of comoving distance, as can be seen in Eq. \eqref{sf}. We have then modified the original software in order to obtain fourth order function derivatives.

\begin{figure*}
\begin{center}
\includegraphics[width=.49\textwidth]{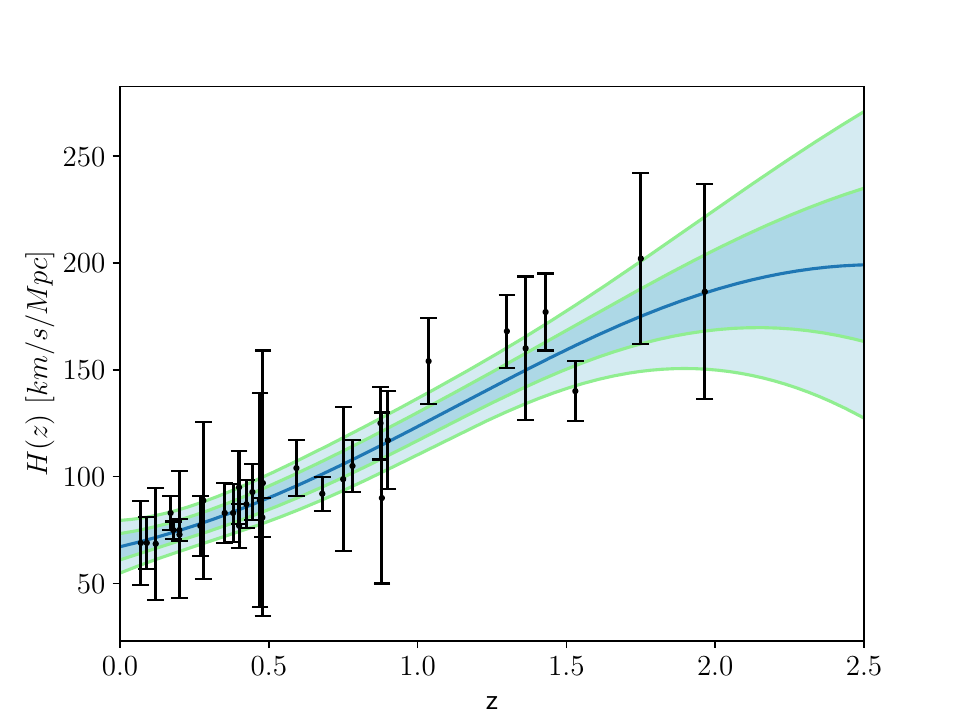}
\includegraphics[width=.49\textwidth]{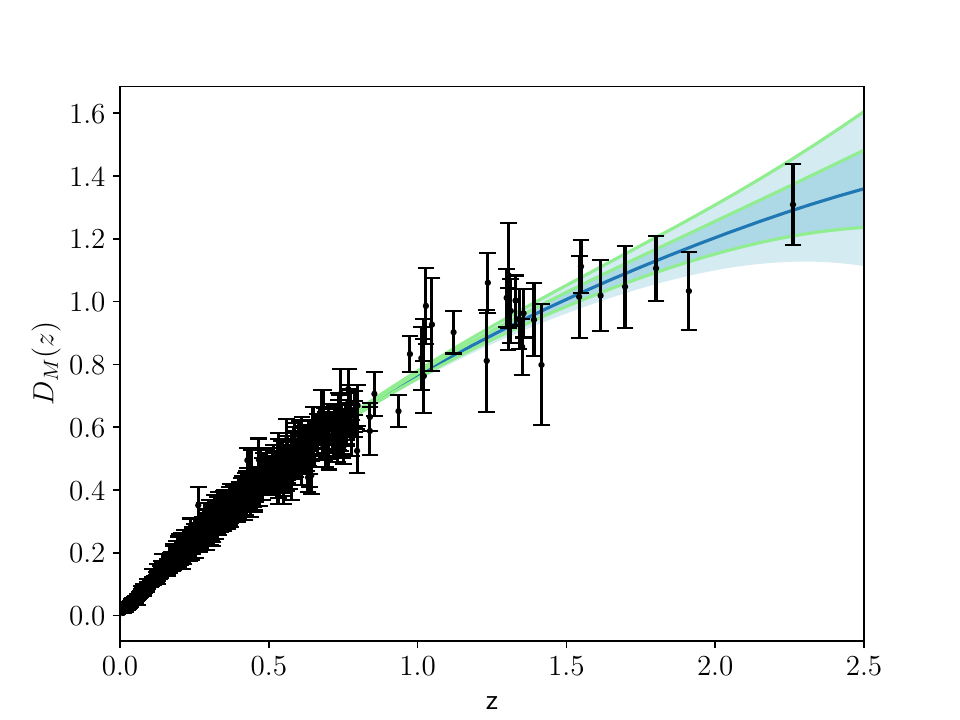}
\end{center}
\caption{GP reconstructions from data. \textbf{a) Left:} Reconstruction of $H(z)$ (in km/s/Mpc) function from 32 $H(z)$ data with covariance. \textbf{b) Right:} Reconstruction of dimensionless $D_M(z)$ from Pantheon$+$ SNe Ia data.}
\label{recdados}
\end{figure*}

Using the GP method, we were able to reconstruct the Hubble parameter $H(z)$ from the cosmic chronometers data, and the $D_M(z)$ using the Pantheon$+$ sample data. These reconstructions are shown in Fig. \ref{recdados}, where the left figure corresponds to the plot of cosmic chronometers data together with the reconstruction of $H(z)$ and the right figure shows the reconstruction and the data of $D_M(z)$ obtained from the Pantheon$+$ sample.

In order to obtain the reconstruction of the kinematic parameters in terms of $D_M(z)$, using Eqs. \eqref{E2DM}, \eqref{qzDc}, \eqref{jzDc} and \eqref{szDc}, however, we need to get a free parameter, $\Omega_k$. This parameter is not obtained from the reconstruction, it must be constrained with the aid of other observations, which can provide a prior on $\Omega_k$.

The prior used was provided by Planck 18 (TT,TE,EE+lowE) (\cite{Planck18}), where $\Omega_k = -0.044\pm0.050, $ at 3$\sigma$ c.l. We chose to work with a 3$\sigma$ prior in order to allow for a more model-independent analysis. With this prior over $(\Omega_k)$, we can obtain the kinematic parameters by sampling this prior and the multivariate Gaussian corresponding to $H(z)$, $D_M(z)$ and their derivatives.

We also performed the reconstruction of the kinematic parameters with the combination of  $H(z)$ data and the Pantheon$+$ sample, in the redshift range where the respective reconstructions were compatible in $1\sigma$. This analysis was performed by weighting the reconstruction obtained for $D_M(z)$ with the reconstruction of $H(z)$, by a MCMC method known as importance sampling. It is the first time that such a combination of GP reconstructions is made in the literature, as far as we know.

For each reconstruction of kinematic parameter below, we also show the constraints obtained from Planck 18 (Plik,TT,TE,EE+lowE+lensing) (\cite{Planck18}), for a flat $\Lambda$CDM model, corresponding to $\Omega_{m} = 0.3153\pm 0.0073$, $H_0=67.36\pm0.54$ km/s/Mpc. Although in the model-independent GP analysis we have chosen to allow for some spatial curvature, we chose to compare with the flat $\Lambda$CDM model, as this is regarded as the standard, cosmic concordance model. Concerning the $H_0$ tension, as Pantheon$+$ alone does not constrain $H_0$ and we intended to combine reconstructions from $H(z)$ and SNe Ia data, we do not reconstruct $H(z)$, but $E(z)$ instead, which is independent of $H_0$. In the end of the section, however, we compare the $H_0$ determinations from $H(z)$, Planck 18 and SH0ES.

\subsection{\texorpdfstring{$E(z)$}{Ez} Reconstruction}
\begin{figure*}
\begin{center}
\includegraphics[width=.49\textwidth]{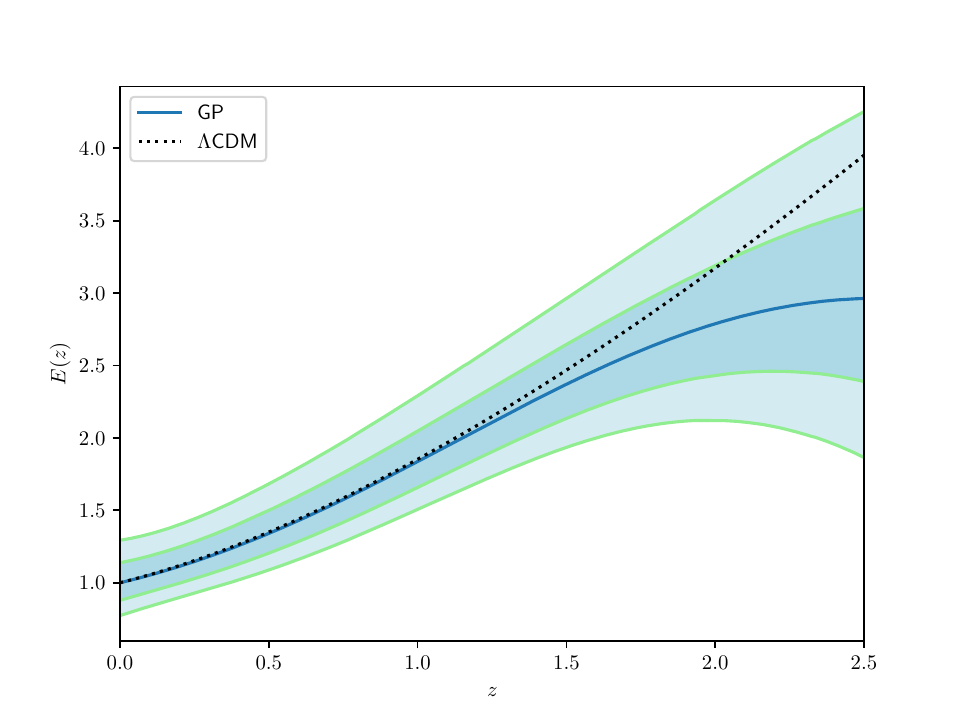}
\includegraphics[width=.49\textwidth]{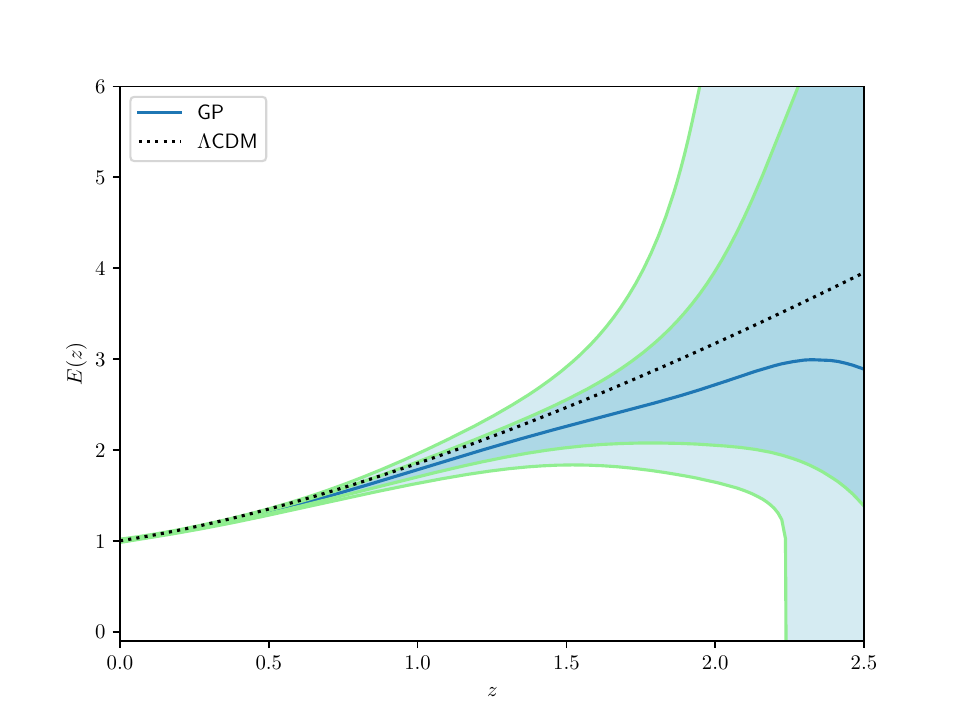}
\includegraphics[width=.49\textwidth]{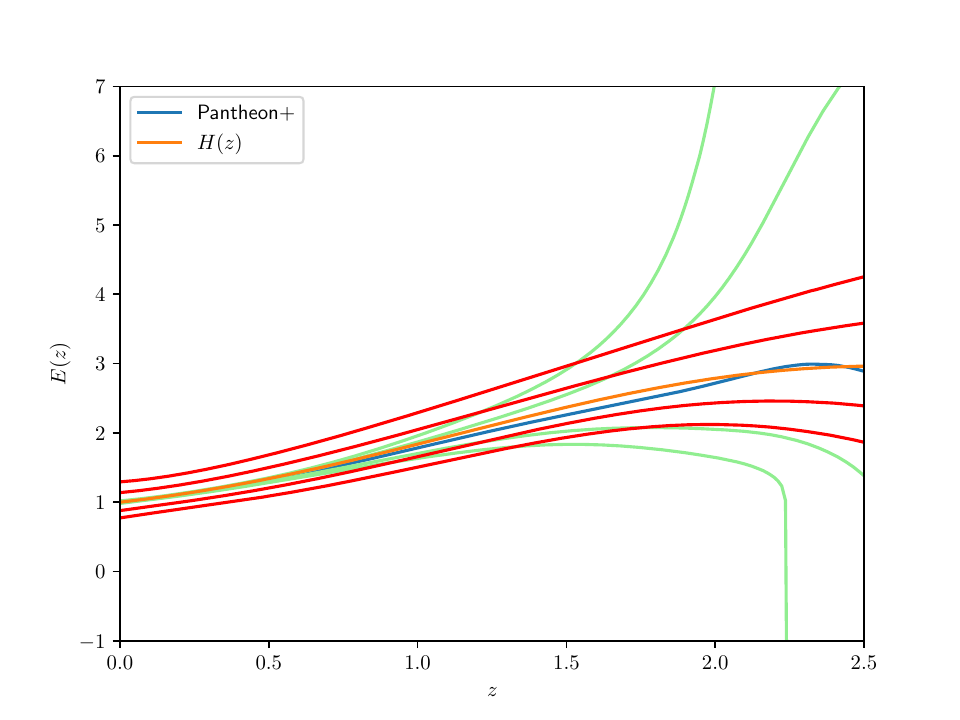}
\includegraphics[width=.49\textwidth]{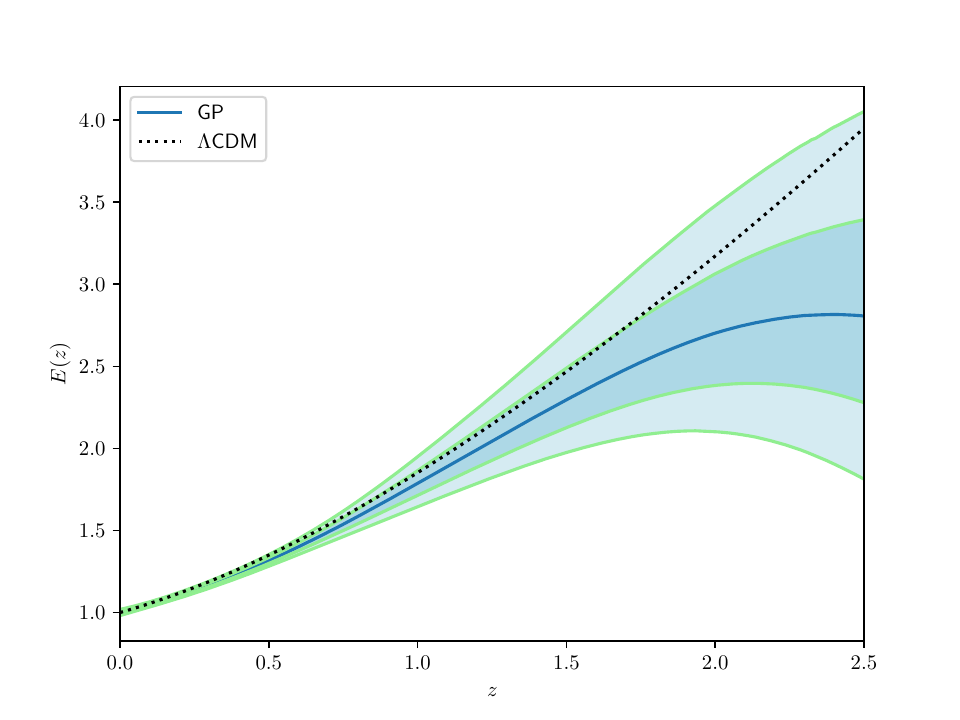}
\end{center}

\caption{$E(z)$ reconstruction. \textbf{a) Top Left:} Reconstruction from $H(z)$ data. \textbf{b) Top Right:} Reconstruction from SNe Ia. \textbf{c) Bottom Left:} Comparison between reconstructions. \textbf{d) Bottom Right:} Joint reconstruction.}
\label{EzRec}
\end{figure*}

The $E(z)$ reconstructions obtained by GP are shown with 1 and $2\sigma$ confidence intervals in Fig. \ref{EzRec}.

In Fig. \ref{EzRec}, at the top left, we show the $E(z)$ reconstruction for the $H(z)$ data. The dashed line corresponds to $E(z)$ for the flat $\Lambda$CDM model, with $\Omega_m$ obtained from Planck 18. The $\Lambda$CDM model prediction is compatible within $1\sigma$ with the GP reconstruction for $z\lesssim2.3$, being compatible within 2$\sigma$ in the remaining interval. The top right figure shows the reconstruction of $E(z)$ made with the {Pantheon$+$} sample data. As we can see, the standard $\Lambda$CDM model is compatible with the GP reconstruction within $1\sigma$ for most part of the redshift interval, being compatible at $2\sigma$ only for an initial interval, \textbf{ $0.1\lesssim z\lesssim0.8$}.
 
In the bottom left figure, we show the superimposed reconstructions in order to see at which redshift interval they are compatible at 1$\sigma$, thus allowing us to safely make a joint analysis. We find that they are compatible within 1$\sigma$ for all the analyzed interval, $0<z<2.5$, thus we proceeded to the full joint analysis, which can be seen in the bottom right figure. As can be seen in Fig. \ref{EzRec}d, {in the interval $0.4\lesssim z\lesssim0.8$ and for $z\gtrsim1.6$}  the standard model is not compatible with GP joint analysis within 1$\sigma$, being compatible at 2$\sigma$, however.
 
\subsection{\texorpdfstring{$q(z)$}{qz} Reconstruction}
\begin{figure*}
\begin{center}
\includegraphics[width=.49\textwidth]{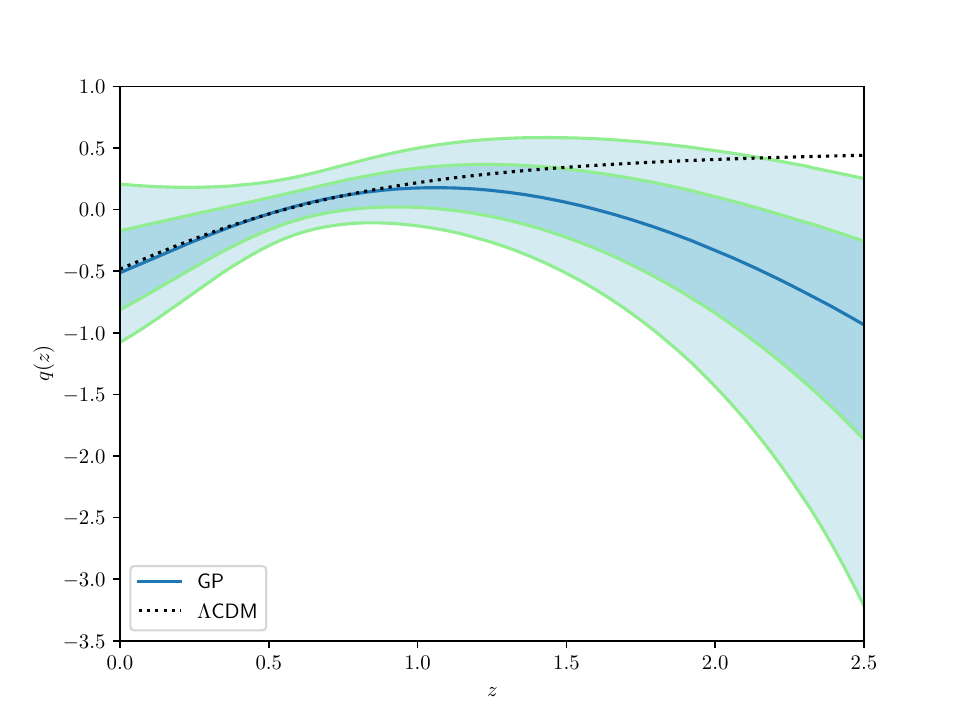}
\includegraphics[width=.49\textwidth]{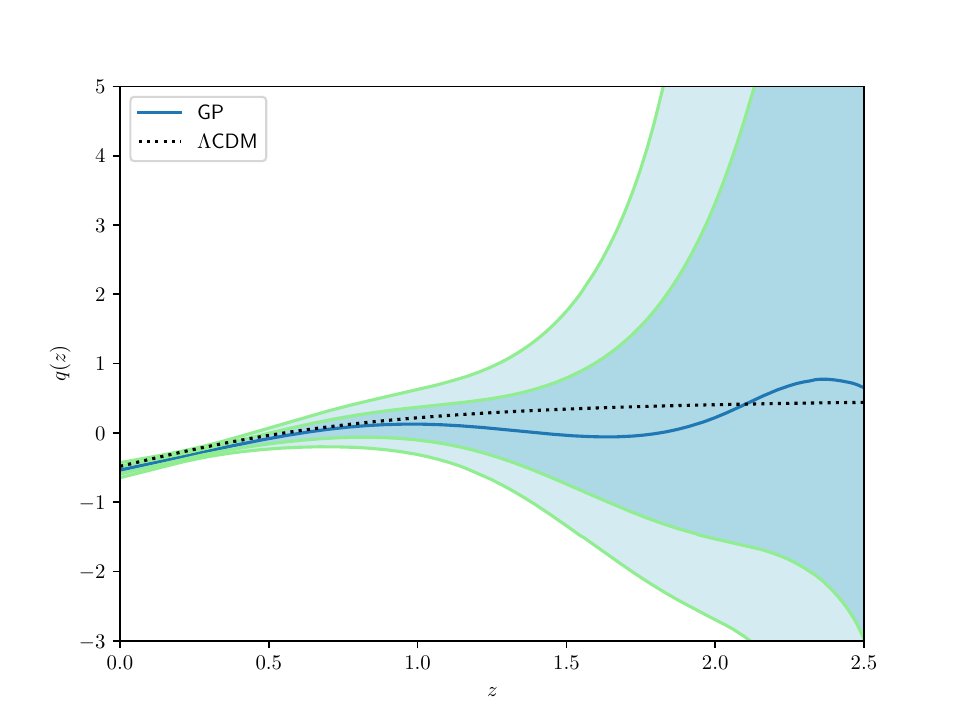}
\includegraphics[width=.49\textwidth]{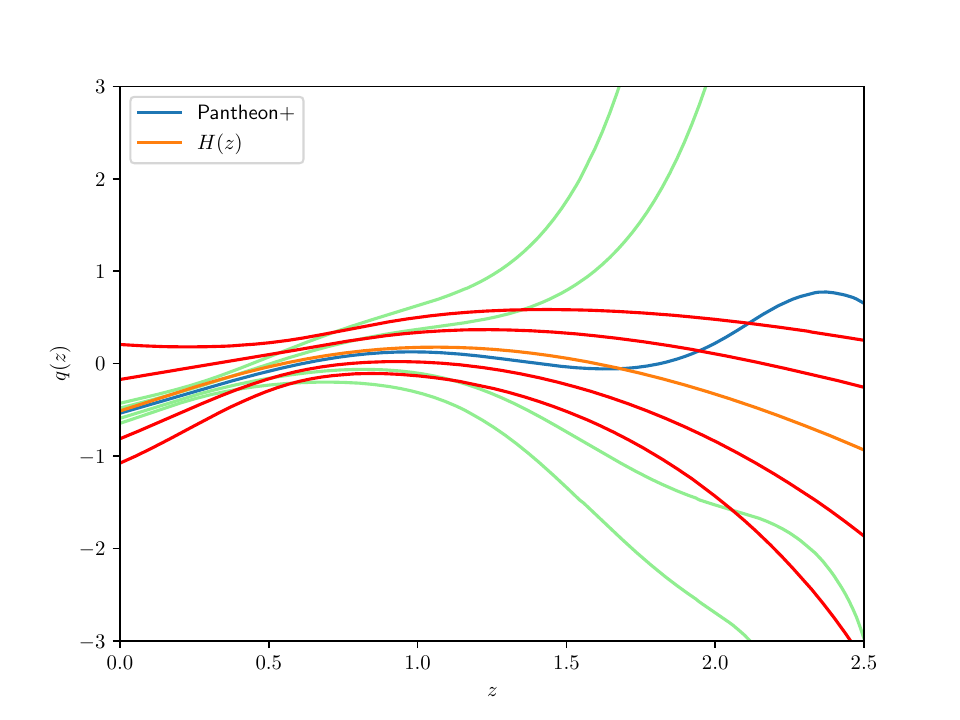}
\includegraphics[width=.49\textwidth]{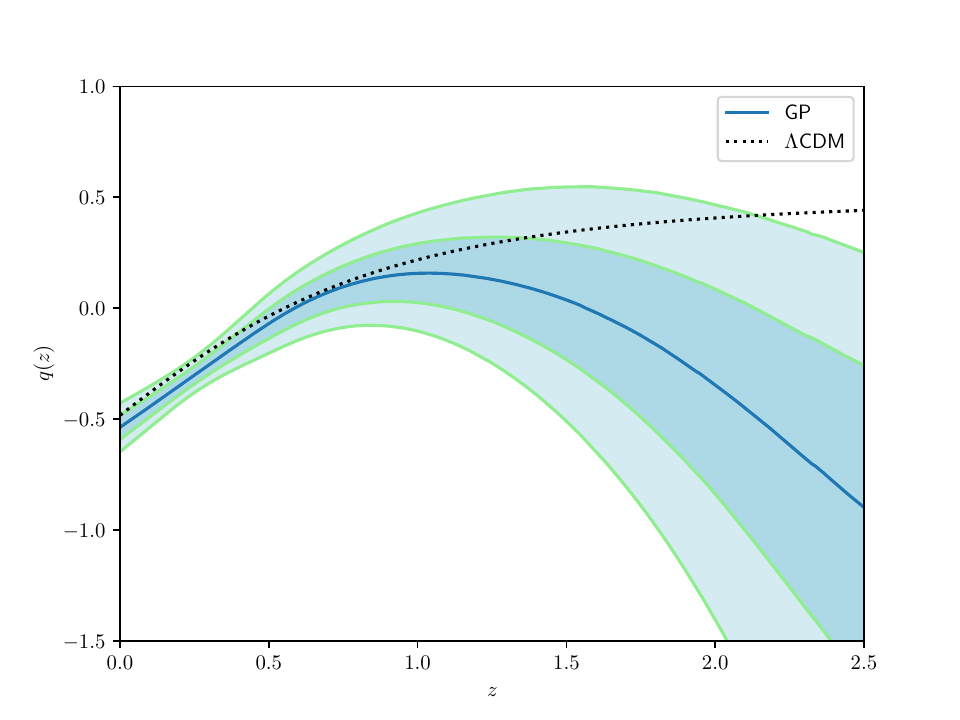}
\end{center}

\caption{$q(z)$ reconstruction. \textbf{a) Top Left:} Reconstruction from $H(z)$ data. \textbf{b) Top Right:} Reconstruction from SNe Ia. \textbf{c) Bottom Left:} Comparison between reconstructions. \textbf{d) Bottom Right:} Joint reconstruction.}
\label{qzRec}
\end{figure*}

In Fig. \ref{qzRec} we show the reconstruction of $q(z)$ with a confidence interval of $2\sigma$.

In Fig. \ref{qzRec}a, we see the $q(z)$ reconstruction from CCs. As one can see, the $\Lambda$CDM model is compatible with the GP reconstruction within 2$\sigma$ for a large interval, up to $z\approx2.3$. For the reconstruction from SNe Ia (Fig. \ref{qzRec}b), however, {compatibility at less than $1\sigma$ is obtained over most of the analyzed range, only at $0\lesssim z\lesssim0.4$ we have compatibility up to $2\sigma$.}

In Fig. \ref{qzRec}c, we superimpose both reconstructions in order to combine them. One may see that they are compatible within 1$\sigma$ {across the analyzed redshift range}, which can be seen on Fig. \ref{qzRec}d. As one may see,{ the $\Lambda$CDM model prediction is within 2$\sigma$ c.l. of the joint reconstruction in $z \lesssim 2.1$.}

\subsection{\texorpdfstring{$j(z)$}{jz} Reconstruction}
\begin{figure*}
\begin{center}
\includegraphics[width=.49\textwidth]{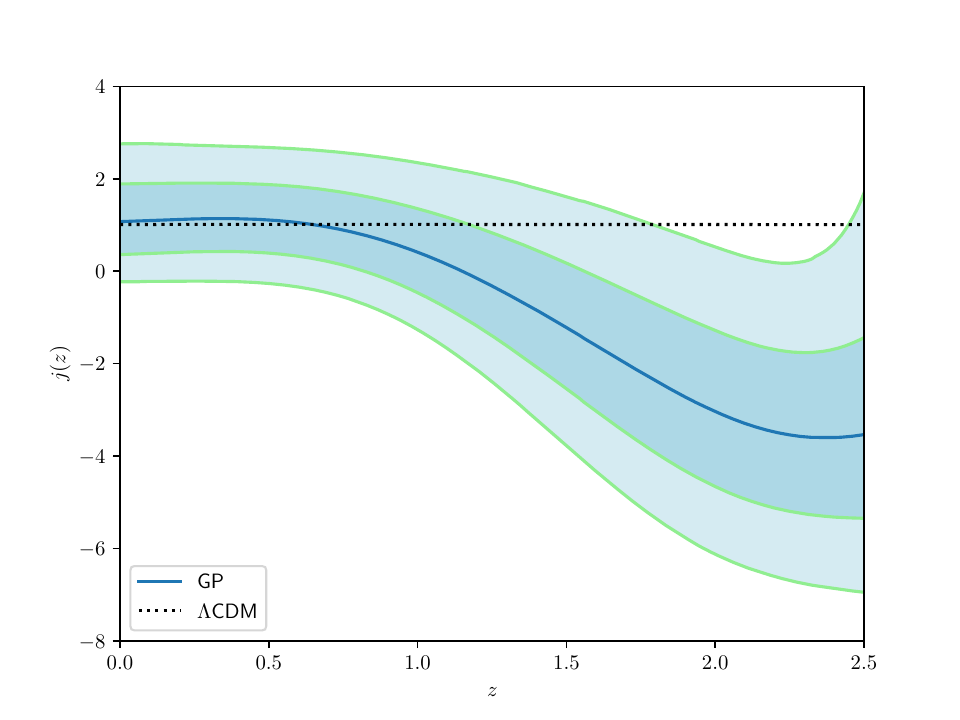}
\includegraphics[width=.49\textwidth]{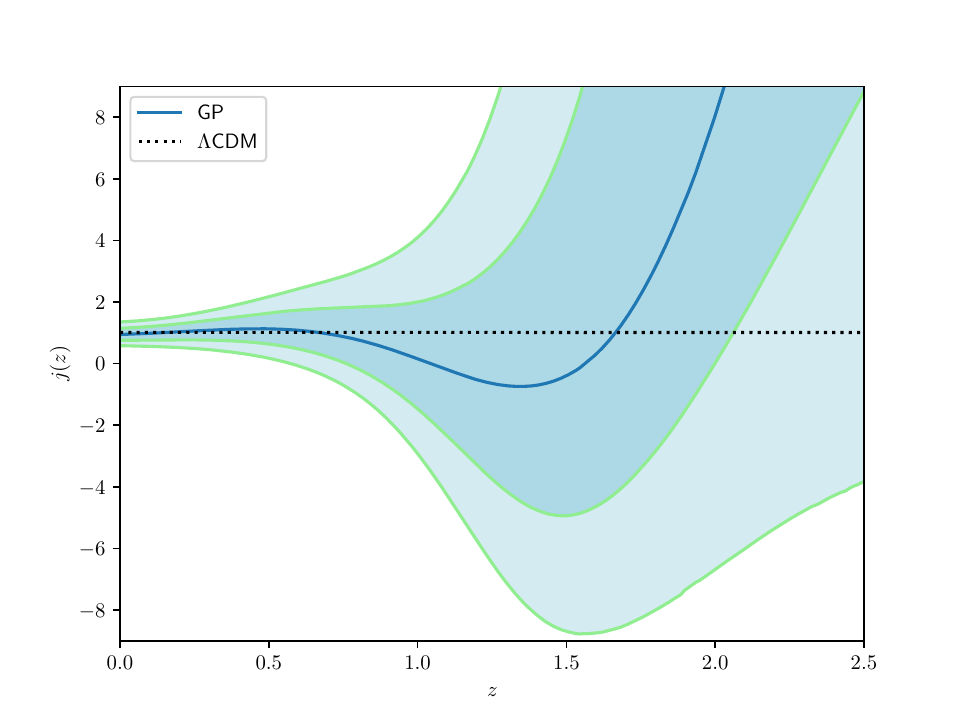}
\includegraphics[width=.49\textwidth]{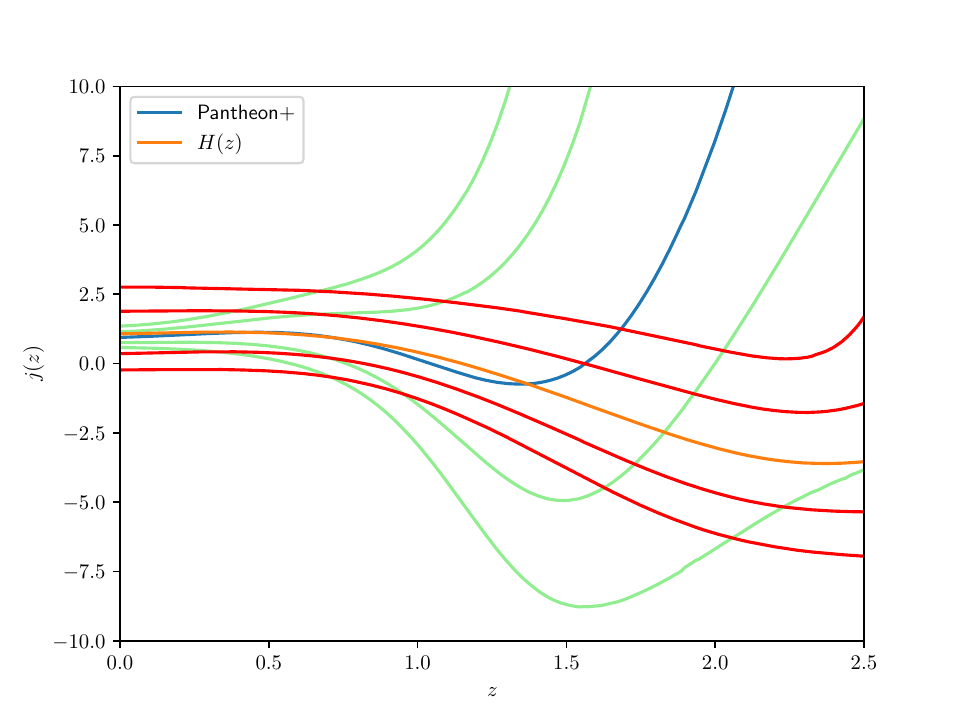}
\includegraphics[width=.49\textwidth]{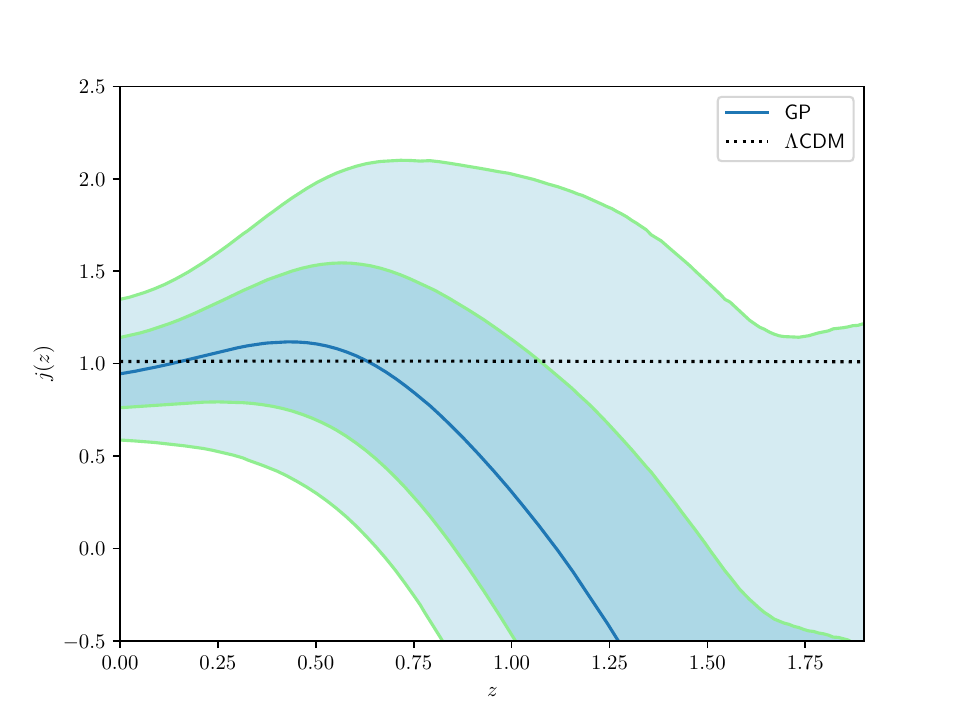}
\end{center}
\caption{$j(z)$ reconstruction. \textbf{a) Top Left:} Reconstruction from $H(z)$ data. \textbf{b) Top Right:} Reconstruction from SNe Ia. \textbf{c) Bottom Left:} Comparison between reconstructions. \textbf{d) Bottom Right:} Joint reconstruction.}
\label{jzRec}
\end{figure*}

We show the reconstruction of $j(z)$ with $2\sigma$ uncertainty in Fig. \ref{jzRec}.

In Fig. \ref{jzRec}a, we may see the jerk reconstruction from CCs. The comparison with the $\Lambda$CDM model here is quite interesting, as the standard, flat $\Lambda$CDM model predict a exact value $j(z)=1$. One may see from this Figure, that only at a high reshift interval, $1.9\lesssim z\lesssim2.4$, $\Lambda$CDM is outside of the 2$\sigma$ interval of the GP reconstruction. For the SNe Ia reconstruction, however, one may see in Fig. \ref{jzRec}b,{ that the $\Lambda$CDM model is initially within 1$\sigma$ until $z\approx2.0$ and then goes into the region of 2$\sigma$.} 


In Fig. \ref{jzRec}c, we have superimposed both reconstructions. One may see that both reconstructions agree { for $z\lesssim 1.9$}, so, in Fig. \ref{jzRec}d, we made the combination of the reconstructions only in this interval. One may see that the $\Lambda$CDM model agrees with the joint reconstruction for all this redshift interval, within 2$\sigma$ c.l.{, this compatibility is less than 1$\sigma$ c.l. in approximately $90\%$ of the redshift interval.}

\subsection{\texorpdfstring{$s(z)$}{sz} Reconstruction}
\begin{figure*}
\begin{center}
\includegraphics[width=.49\textwidth]{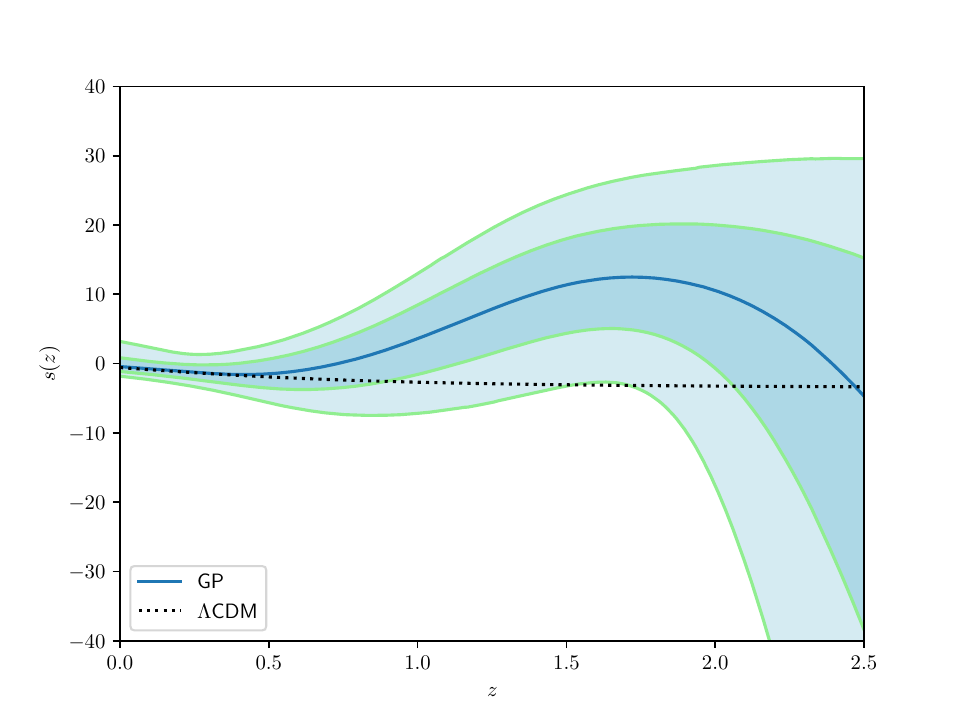}
\includegraphics[width=.49\textwidth]{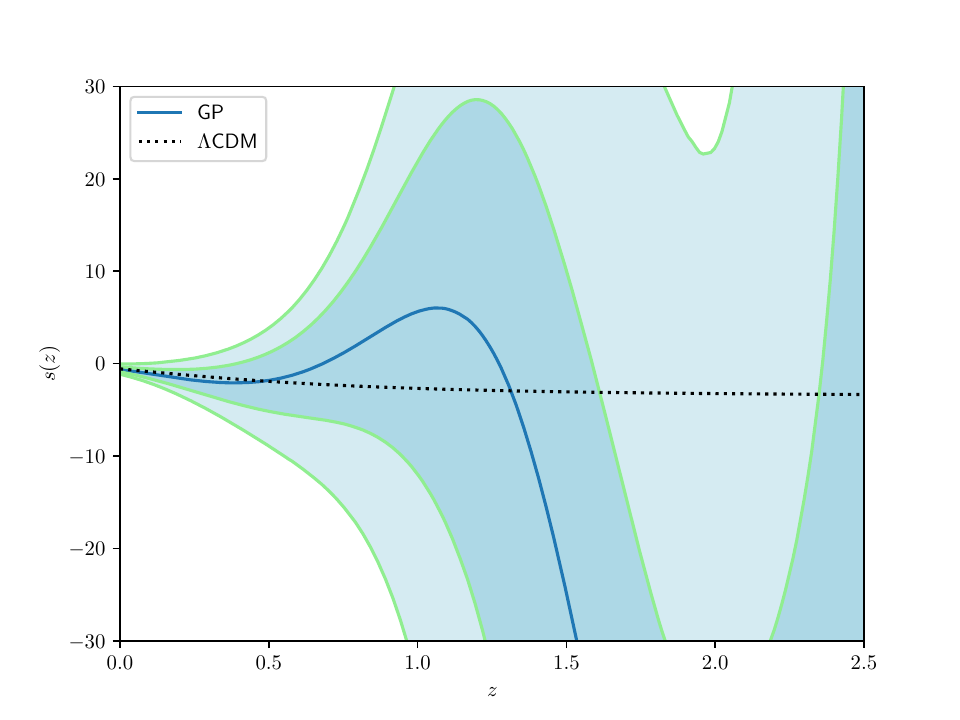}
\includegraphics[width=.49\textwidth]{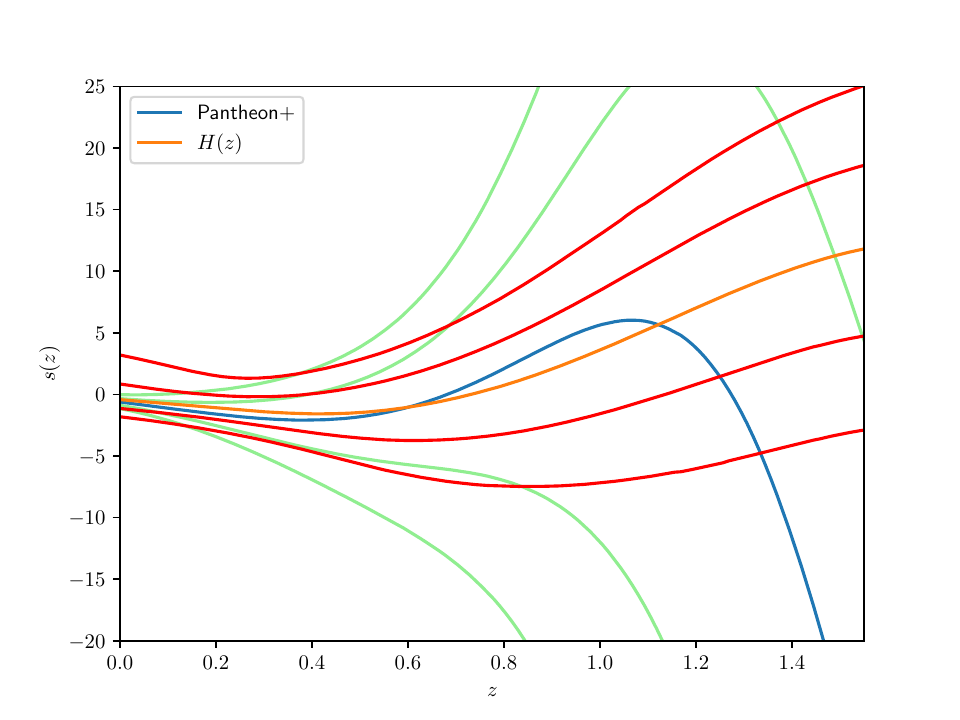}
\includegraphics[width=.49\textwidth]{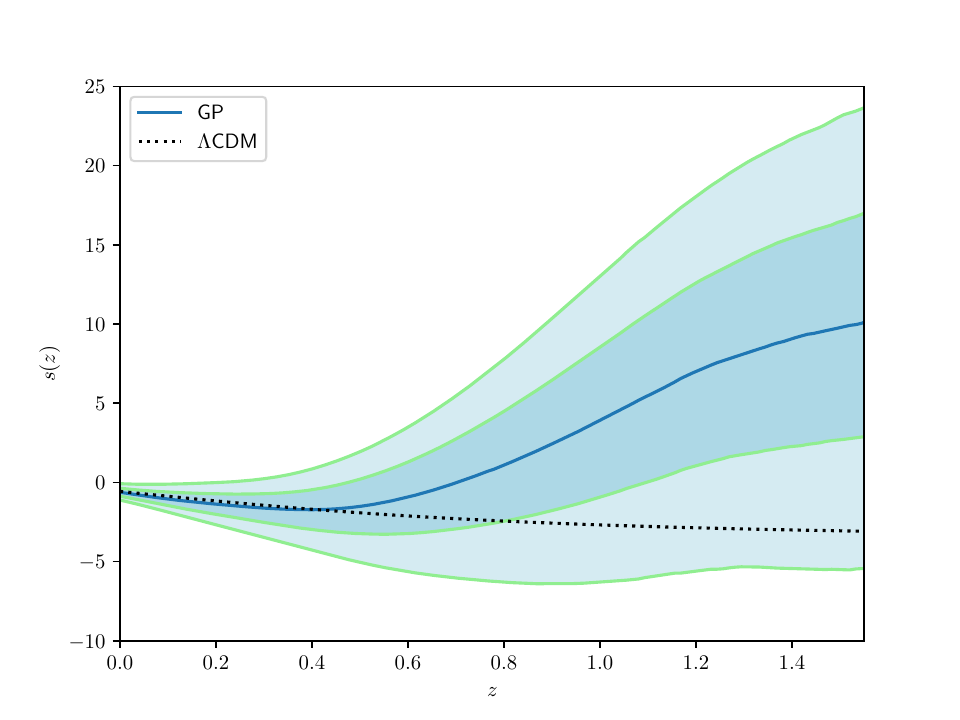}
\end{center}
\caption{$s(z)$ reconstruction. \textbf{a) Top Left:} Reconstruction from $H(z)$ data. \textbf{b) Top Right:} Reconstruction from SNe Ia. \textbf{c) Bottom Left:} Comparison between reconstructions. \textbf{d) Bottom Right:} Joint reconstruction.}
\label{szRec}
\end{figure*}

The reconstruction of $s(z)$ obtained with GP is shown in Fig. \ref{szRec}.

As can be seen on Figs. \ref{szRec}a and \ref{szRec}b, the $\Lambda$CDM model is compatible within 2$\sigma$ with both reconstructions. In Fig. \ref{szRec}c, one may see that both reconstructions agree within 1$\sigma$ only for a small interval, $z\lesssim 1.6$. The joint reconstruction in this interval can be seen on Fig. \ref{szRec}d, where can be seen that the $\Lambda$CDM model is compatible with it for this whole interval {in until 2$\sigma$ c.l..} 



\subsection{Current values of kinematic parameters from Gaussian Processes}
As one could see above, the kinematic parameters were better constrained at lower redshifts. So, in this subsection, we focus in the constraints for the current values of the kinematic parameters, which have great interest in the literature. In Table \ref{tab1}, we show the values of the kinematic parameters today in $z=0$, which we obtained through the GP reconstructions. We also show the constraints obtained from Planck 18 (\cite{Planck18}) for a flat $\Lambda$CDM model, corresponding to $\Omega_{m} = 0.315\pm 0.007$. We have compatibility within $1\sigma$ c.l. for the $\Lambda$CDM model with our results. The value of $H_0$ that we obtain with GP is also compatible in $1\sigma$ with the value $H_0 = 73.2 \pm 1.7\pm 2.4$ km/s/Mpc found by SH0ES (\cite{Riess:2016jrr}).

We should also compare these values with the ones obtained of kinematic parametrizations in \cite{BenndorfEtAl22}. In their best parametrization, $j(z)=j_0$, they have obtained $H_0=68.8\pm1.9$ km/s/Mpc,  $q_0=-0.578\pm0.067$, $j_0=1.15\pm0.28$ and $s_0=-0.255^{+0.094}_{-0.21}$ at 1$\sigma$ c.l. This result is from Pantheon+31 $H(z)$ data from \cite{MaganaEtAl17}, without systematics. We may see that this is compatible within 1$\sigma$ with our results in Table \ref{tab1}. However, while the $H_0$ value was compatible with SH0ES only at 2$\sigma$ c.l., we now have compatibility with SH0ES at 1$\sigma$ c.l. thanks to taking the systematics into account. {We can also compare these results with, for instance, \cite{Mukherjee21}, where they have reconstructed $H(z)$, $q(z)$ and $j(z)$, with GP, using CCs and SNe Ia (Pantheon). They have found from CC alone, $H_0=71.2\pm4.6$ km/s/Mpc, which is compatible with our result within 1 $\sigma$ c.l. The other results they have found, namely, $q_0$, $j_0$ from CC, SNe Ia and CC+SNe Ia data (Tab. 5 of \cite{Mukherjee21}), are all compatible with our results of Tab. \ref{tab1}. It is important that they have not considered $H(z)$ systematics, they have used the older Pantheon (\cite{pantheon}) samples and they have used error propagation in order to obtain these uncertainties estimates.}

\begin{table*}
\begin{tabular}{|c|c|c|c|c|}
\hline
\multicolumn{1}{|c|}{Parameter}&
\multicolumn{3}{|c|}{Gaussian Processes Reconstruction}&\multicolumn{1}{|c|}{$\Lambda$CDM Model} \\\cline{2-5}
 & $H(z)$ & SNe Ia & Joint& Planck 18  \\\hline
  $H_0$ (km/s/Mpc)  & $67.2 \pm 6.2\pm 12.3$ &  &  & $67.4 \pm 0.5\pm 1.0$\\\hline
 $q(z=0)$ & $-0.51^{+0.34+0.72}_{-0.30-0.56}$ & $-0.54^{+0.05+0.11}_{-0.06-0.11}$ & $-0.54^{+0.06+0.11}_{-0.05-0.11}$ & $-0.528\pm 0.011\pm 0.021$ \\\hline
  $j(z=0)$ & $1.07^{+0.82+1.68}_{-0.71-1.30}$ & $0.94^{+0.20+0.41}_{-0.18-0.36}$ & $0.94^{+0.20+0.40}_{-0.18-0.36}$&  $1$\\\hline
$s(z=0)$&$-0.43^{+1.3+3.6}_{-0.71-1.4}$ & $-0.62^{+0.28+0.59}_{-0.27-0.53}$ & $-0.62^{+0.26+0.54}_{-0.25-0.49}$&$-0.418\pm 0.032 \pm 0.064$ \\\hline
\end{tabular}
\caption{Current values of kinematic parameters ($z=0$), obtained with GP.}
    \label{tab1}
\end{table*}

\begin{figure*}
\begin{center}
\includegraphics[width=.49\textwidth]{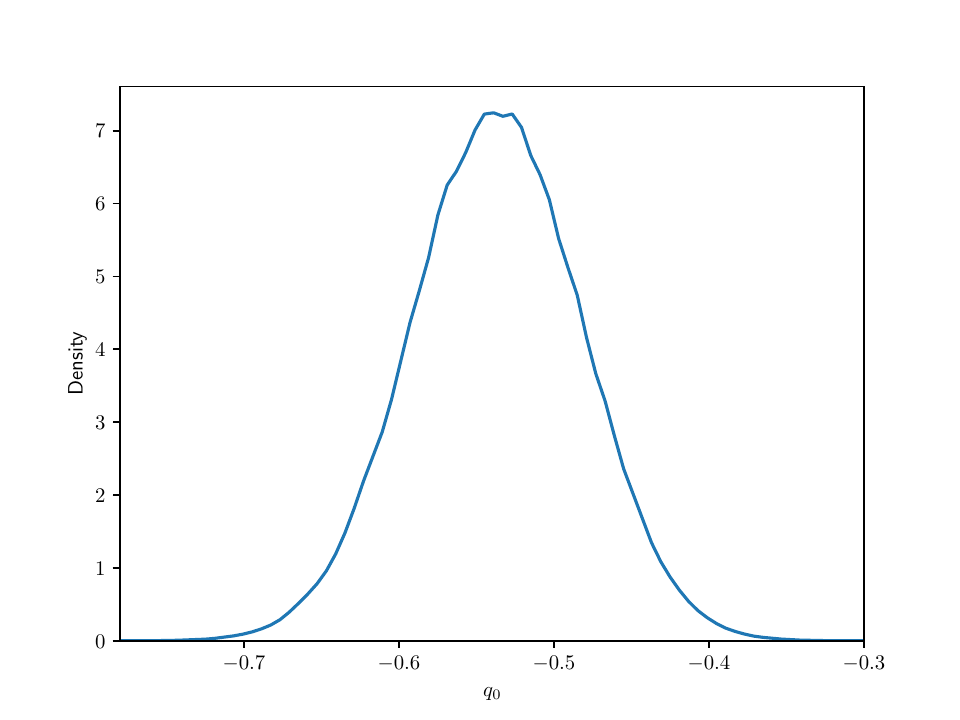}\\
\includegraphics[width=.49\textwidth]{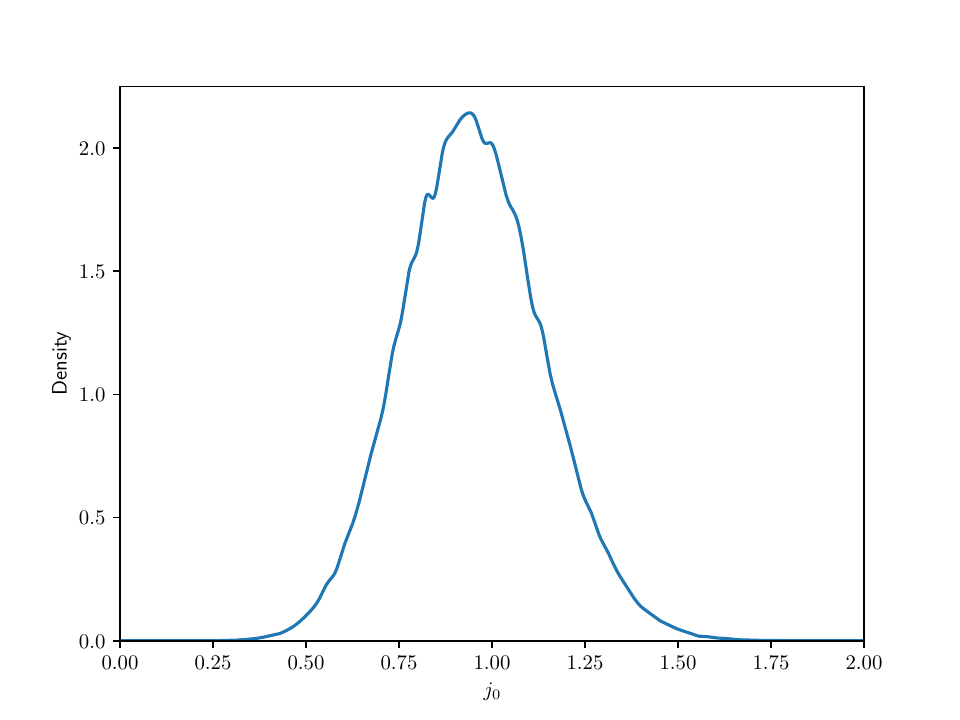}
\includegraphics[width=.49\textwidth]{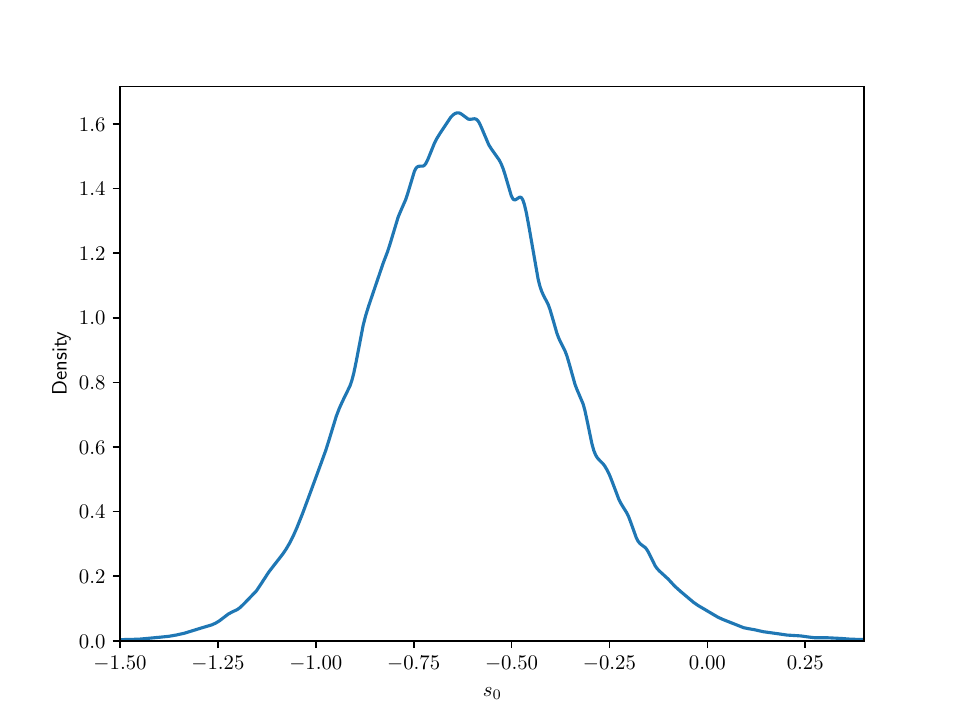}
\end{center}
\caption{Current values of the kinematic parameters from the GP joint reconstructions.}
\label{current}
\end{figure*}

In Fig. \ref{current}, we show the joint distributions found for the current values of the kinematic parameters $q_0$, $j_0$ and $s_0$. We do not show the $H_0$ distribution here because it comes only from $H(z)$ data and it is a Normal distribution as expected from Gaussian Processes.

\section{Conclusions}
We have reconstructed the evolution of four kinematic parameters, namely, $E(z)$, $q(z)$, $j(z)$ and $s(z)$, from Pantheon$+$ and $H(z)$ data. For the first time, we have used $H(z)$ data with systematical errors in this kind of analysis. In order to obtain model-independent reconstructions, we have allowed for a prior in the spatial curvature, coming from Planck 18 analysis. Differently from earlier analyses, $H_0$ obtained from $H(z)$ data is now compatible within 1$\sigma$ both with Planck 18 and SH0ES, thereby not favouring any of the poles of the $H_0$ tension. We have made a combination of the reconstructions in each case, for the redshift intervals where it was safe to make it. For all reconstructions, the so-called cosmic concordance flat $\Lambda$CDM model is compatible within 2$\sigma$ in the analyzed redshift intervals. A general tendency is that SNe Ia constrain better at lower redshifts, while $H(z)$ data constrain better at higher redshifts, thereby showing the importance of such a combination. Both reconstructions agree at lower redshifts and lower derivatives, while being less compatible at higher redshifts and derivatives.

Other possibilities, as including more data and reconstructing other kinematic parameters can be explored in a forthcoming issue.

\section*{Acknowledgements}
The authors would like to thank the referees for
their constructive suggestions which improved the original version of our paper. The  authors are grateful  to CAPES - Coordena\c{c}\~ao de Aperfei\c{c}oamento de Pessoal de N\'ivel Superior and to CNPq - Conselho Nacional de Desenvolvimento Científico e Tecnológico  (Brazilian Research Agencies). JFJ acknowledges Felipe Andrade-Oliveira for helpful discussions.

\section*{Data Availability}

The data underlying this article are freely available in \cite{pantheonplus} and \cite{MorescoEtAl22}.



\bibliographystyle{mnras}
\bibliography{references} 

@article{Kjerrgren:2021zuo,
    author = "Kjerrgren, Anders Ahlstrom and Mortsell, Edvard",
    title = "{On the use of galaxies as clocks and the universal expansion}",
    eprint = "2106.11317",
    archivePrefix = "arXiv",
    primaryClass = "astro-ph.CO",
    doi = "10.1093/mnras/stac1978",
    journal = "Mon. Not. Roy. Astron. Soc.",
    volume = "518",
    number = "1",
    pages = "585--591",
    year = "2022"
}

@article{Planck18,
    author = "Aghanim, N. and others",
    collaboration = "Planck",
    title = "{Planck 2018 results. VI. Cosmological parameters}",
    eprint = "1807.06209",
    archivePrefix = "arXiv",
    primaryClass = "astro-ph.CO",
    doi = "10.1051/0004-6361/201833910",
    journal = "Astron. Astrophys.",
    volume = "641",
    pages = "A6",
    year = "2020",
    note = "[Erratum: Astron.Astrophys. 652, C4 (2021)]"
}

@article{pantheonplus,
    author = "Brout, Dillon and others",
    title = "{The Pantheon+ Analysis: Cosmological Constraints}",
    eprint = "2202.04077",
    archivePrefix = "arXiv",
    primaryClass = "astro-ph.CO",
    doi = "10.3847/1538-4357/ac8e04",
    journal = "Astrophys. J.",
    volume = "938",
    number = "2",
    pages = "110",
    year = "2022"
}

@article{pantheon,
    author = "Scolnic, D. M. and others",
    collaboration = "Pan-STARRS1",
    title = "{The Complete Light-curve Sample of Spectroscopically Confirmed SNe Ia from Pan-STARRS1 and Cosmological Constraints from the Combined Pantheon Sample}",
    eprint = "",
    archivePrefix = "arXiv",
    primaryClass = "astro-ph.CO",
    doi = "",
    journal = "Astrophys. J.",
    volume = "859",
    number = "2",
    pages = "101",
    year = "2018"
}

@article{SeikelEtAl12a,
    author = "Seikel, Marina and Clarkson, Chris and Smith, Mathew",
    title = "{Reconstruction of dark energy and expansion dynamics using Gaussian processes}",
    eprint = "1204.2832",
    archivePrefix = "arXiv",
    primaryClass = "astro-ph.CO",
    doi = "10.1088/1475-7516/2012/06/036",
    journal = "JCAP",
    volume = "06",
    pages = "036",
    year = "2012"
}

@article{MaganaEtAl17,
    author = "Magana, Juan and Amante, Mario H. and Garcia-Aspeitia, Miguel A. and Motta, V.",
    title = "{The Cardassian expansion revisited: constraints from updated Hubble parameter measurements and type Ia supernova data}",
    eprint = "1706.09848",
    archivePrefix = "arXiv",
    primaryClass = "astro-ph.CO",
    doi = "10.1093/mnras/sty260",
    journal = "Mon. Not. Roy. Astron. Soc.",
    volume = "476",
    number = "1",
    pages = "1036--1049",
    year = "2018"
}

@article{Bull2016,
    author = "Bull, Philip and others",
    title = "{Beyond $\Lambda$CDM: Problems, solutions, and the road ahead}",
    eprint = "1512.05356",
    archivePrefix = "arXiv",
    primaryClass = "astro-ph.CO",
    doi = "10.1016/j.dark.2016.02.001",
    journal = "Phys. Dark Univ.",
    volume = "12",
    pages = "56--99",
    year = "2016"
}

@article{GongBo,
    author = "Zhao, Gong-Bo and others",
    title = "{Dynamical dark energy in light of the latest observations}",
    eprint = "1701.08165",
    archivePrefix = "arXiv",
    primaryClass = "astro-ph.CO",
    doi = "10.1038/s41550-017-0216-z",
    journal = "Nature Astron.",
    volume = "1",
    number = "9",
    pages = "627--632",
    year = "2017"
}

@article{marttens2020,
    author = "von Marttens, Rodrigo and Lombriser, Lucas and Kunz, Martin and Marra, Valerio and Casarini, Luciano and Alcaniz, Jailson",
    title = "{Dark degeneracy I: Dynamical or interacting dark energy?}",
    eprint = "1911.02618",
    archivePrefix = "arXiv",
    primaryClass = "astro-ph.CO",
    doi = "10.1016/j.dark.2020.100490",
    journal = "Phys. Dark Univ.",
    volume = "28",
    pages = "100490",
    year = "2020"
}

@article{SF,
    author = "Pereira, S. H. and Jesus, J. F.",
    title = "{Can Dark Matter Decay in Dark Energy?}",
    eprint = "0811.0099",
    archivePrefix = "arXiv",
    primaryClass = "astro-ph",
    doi = "10.1103/PhysRevD.79.043517",
    journal = "Phys. Rev. D",
    volume = "79",
    pages = "043517",
    year = "2009"
}

@article{Majerotto2009,
    author = "Majerotto, Elisabetta and Valiviita, Jussi and Maartens, Roy",
    title = "{Adiabatic initial conditions for perturbations in interacting dark energy models}",
    eprint = "0907.4981",
    archivePrefix = "arXiv",
    primaryClass = "astro-ph.CO",
    doi = "10.1111/j.1365-2966.2009.16140.x",
    journal = "Mon. Not. Roy. Astron. Soc.",
    volume = "402",
    pages = "2344--2354",
    year = "2010"
}

@article{Valiviita2010,
    author = "Valiviita, Jussi and Maartens, Roy and Majerotto, Elisabetta",
    title = "{Observational constraints on an interacting dark energy model}",
    eprint = "0907.4987",
    archivePrefix = "arXiv",
    primaryClass = "astro-ph.CO",
    doi = "10.1111/j.1365-2966.2009.16115.x",
    journal = "Mon. Not. Roy. Astron. Soc.",
    volume = "402",
    pages = "2355--2368",
    year = "2010"
}

@article{Chimento2010,
    author = "Chimento, Luis P.",
    title = "{Linear and nonlinear interactions in the dark sector}",
    eprint = "0911.5687",
    archivePrefix = "arXiv",
    primaryClass = "astro-ph.CO",
    doi = "10.1103/PhysRevD.81.043525",
    journal = "Phys. Rev. D",
    volume = "81",
    pages = "043525",
    year = "2010"
}

@article{Cai2010,
    author = "Cai, Rong-Gen and Su, Qiping",
    title = "{On the Dark Sector Interactions}",
    eprint = "0912.1943",
    archivePrefix = "arXiv",
    primaryClass = "astro-ph.CO",
    doi = "10.1103/PhysRevD.81.103514",
    journal = "Phys. Rev. D",
    volume = "81",
    pages = "103514",
    year = "2010"
}

@article{Sun2012,
    author = "Sun, Cheng-Yi and Yue, Rui-Hong",
    title = "{New Interaction between Dark Energy and Dark Matter Changes Sign during Cosmological Evolution}",
    eprint = "1009.1214",
    archivePrefix = "arXiv",
    primaryClass = "gr-qc",
    doi = "10.1103/PhysRevD.85.043010",
    journal = "Phys. Rev. D",
    volume = "85",
    pages = "043010",
    year = "2012"
}

@article{Pourtsidou2013,
    author = "Pourtsidou, A. and Skordis, C. and Copeland, E. J.",
    title = "{Models of dark matter coupled to dark energy}",
    eprint = "1307.0458",
    archivePrefix = "arXiv",
    primaryClass = "astro-ph.CO",
    doi = "10.1103/PhysRevD.88.083505",
    journal = "Phys. Rev. D",
    volume = "88",
    number = "8",
    pages = "083505",
    year = "2013"
}

@article{Salvatelli2014,
    author = "Salvatelli, Valentina and Said, Najla and Bruni, Marco and Melchiorri, Alessandro and Wands, David",
    title = "{Indications of a late-time interaction in the dark sector}",
    eprint = "1406.7297",
    archivePrefix = "arXiv",
    primaryClass = "astro-ph.CO",
    doi = "10.1103/PhysRevLett.113.181301",
    journal = "Phys. Rev. Lett.",
    volume = "113",
    number = "18",
    pages = "181301",
    year = "2014"
}

@article{Li2014,
    author = "Li, Yun-He and Zhang, Xin",
    title = "{Large-scale stable interacting dark energy model: Cosmological perturbations and observational constraints}",
    eprint = "1312.6328",
    archivePrefix = "arXiv",
    primaryClass = "astro-ph.CO",
    doi = "10.1103/PhysRevD.89.083009",
    journal = "Phys. Rev. D",
    volume = "89",
    number = "8",
    pages = "083009",
    year = "2014"
}

@article{Skordis2015,
    author = "Skordis, C. and Pourtsidou, A. and Copeland, E. J.",
    title = "{Parametrized post-Friedmannian framework for interacting dark energy theories}",
    eprint = "1502.07297",
    archivePrefix = "arXiv",
    primaryClass = "astro-ph.CO",
    doi = "10.1103/PhysRevD.91.083537",
    journal = "Phys. Rev. D",
    volume = "91",
    number = "8",
    pages = "083537",
    year = "2015"
}

@article{Jimenez2016,
    author = "Beltr\'an Jim\'enez, Jose and Rubiera-Garcia, Diego and S\'aez-G\'omez, Diego and Salzano, Vincenzo",
    title = "{Cosmological future singularities in interacting dark energy models}",
    eprint = "1607.06389",
    archivePrefix = "arXiv",
    primaryClass = "gr-qc",
    doi = "10.1103/PhysRevD.94.123520",
    journal = "Phys. Rev. D",
    volume = "94",
    number = "12",
    pages = "123520",
    year = "2016"
}

@article{Valent2020,
    author = "G\'omez-Valent, Adri\`a and Pettorino, Valeria and Amendola, Luca",
    title = "{Update on coupled dark energy and the $H_0$ tension}",
    eprint = "2004.00610",
    archivePrefix = "arXiv",
    primaryClass = "astro-ph.CO",
    doi = "10.1103/PhysRevD.101.123513",
    journal = "Phys. Rev. D",
    volume = "101",
    number = "12",
    pages = "123513",
    year = "2020"
}

@article{kine1,
    author = "Visser, Matt",
    title = "{Jerk and the cosmological equation of state}",
    eprint = "gr-qc/0309109",
    archivePrefix = "arXiv",
    doi = "10.1088/0264-9381/21/11/006",
    journal = "Class. Quant. Grav.",
    volume = "21",
    pages = "2603--2616",
    year = "2004"
}

@article{kine2,
    author = "Visser, Matt",
    editor = "McClelland, D. E. and Scott, S. M.",
    title = "{Cosmography: Cosmology without the Einstein equations}",
    eprint = "gr-qc/0411131",
    archivePrefix = "arXiv",
    doi = "10.1007/s10714-005-0134-8",
    journal = "Gen. Rel. Grav.",
    volume = "37",
    pages = "1541--1548",
    year = "2005"
}

@article{kine3,
    author = "Shapiro, Charles and Turner, Michael S.",
    title = "{What do we really know about cosmic acceleration?}",
    eprint = "astro-ph/0512586",
    archivePrefix = "arXiv",
    doi = "10.1086/506470",
    journal = "Astrophys. J.",
    volume = "649",
    pages = "563--569",
    year = "2006"
}

@article{kine4,
    author = "Blandford, R. D. and Amin, Mustafa A. and Baltz, E. A. and Mandel, K. and Marshall, P. J.",
    title = "{Cosmokinetics}",
    eprint = "astro-ph/0408279",
    archivePrefix = "arXiv",
    reportNumber = "SLAC-PUB-10814",
    journal = "ASP Conf. Ser.",
    volume = "339",
    pages = "27",
    year = "2005"
}

@article{kine5,
    author = "Elgaroy, Oystein and Multamaki, Tuomas",
    title = "{Bayesian analysis of friedmannless cosmologies}",
    eprint = "astro-ph/0603053",
    archivePrefix = "arXiv",
    doi = "10.1088/1475-7516/2006/09/002",
    journal = "JCAP",
    volume = "09",
    pages = "002",
    year = "2006"
}

@article{kine6,
    author = "Rapetti, David and Allen, Steven W. and Amin, Mustafa A. and Blandford, Roger D.",
    title = "{A kinematical approach to dark energy studies}",
    eprint = "astro-ph/0605683",
    archivePrefix = "arXiv",
    reportNumber = "SLAC-PUB-11882",
    doi = "10.1111/j.1365-2966.2006.11419.x",
    journal = "Mon. Not. Roy. Astron. Soc.",
    volume = "375",
    pages = "1510--1520",
    year = "2007"
}

@article{kine7,
    author = "Riess, Adam G. and others",
    title = "{New Hubble Space Telescope Discoveries of Type Ia Supernovae at z\ensuremath{>}=1: Narrowing Constraints on the Early Behavior of Dark Energy}",
    eprint = "astro-ph/0611572",
    archivePrefix = "arXiv",
    reportNumber = "46455850950",
    doi = "10.1086/510378",
    journal = "Astrophys. J.",
    volume = "659",
    pages = "98--121",
    year = "2007"
}

@article{rezaei2021,
    author = "Rezaei, Mehdi and Sol\`a Peracaula, Joan and Malekjani, Mohammad",
    title = "{Cosmographic approach to Running Vacuum dark energy models: new constraints using BAOs and Hubble diagrams at higher redshifts}",
    eprint = "2108.06255",
    archivePrefix = "arXiv",
    primaryClass = "astro-ph.CO",
    doi = "10.1093/mnras/stab3117",
    journal = "Mon. Not. Roy. Astron. Soc.",
    volume = "509",
    number = "2",
    pages = "2593--2608",
    year = "2021"
}

@article{mehrabi2021,
    author = "Mehrabi, Ahmad and Rezaei, Mehdi",
    title = "{Cosmographic Parameters in Model-independent Approaches}",
    eprint = "2110.14950",
    archivePrefix = "arXiv",
    primaryClass = "astro-ph.CO",
    doi = "10.3847/1538-4357/ac2fff",
    journal = "Astrophys. J.",
    volume = "923",
    number = "2",
    pages = "274",
    year = "2021"
}

@article{velasques2021,
    author = "Velasquez-Toribio, A. M. and Fabris, J\'ulio C.",
    title = "{Constraints on Cosmographic Functions of Cosmic Chronometers Data Using Gaussian Processes}",
    eprint = "2104.07356",
    archivePrefix = "arXiv",
    primaryClass = "astro-ph.CO",
    doi = "10.1007/s13538-022-01113-8",
    journal = "Braz. J. Phys.",
    volume = "52",
    number = "4",
    pages = "115",
    year = "2022"
}

@article{lobo2020,
    author = "Lobo, Francisco S. N. and Mimoso, Jose Pedro and Visser, Matt",
    title = "{Cosmographic analysis of redshift drift}",
    eprint = "2001.11964",
    archivePrefix = "arXiv",
    primaryClass = "gr-qc",
    doi = "10.1088/1475-7516/2020/04/043",
    journal = "JCAP",
    volume = "04",
    pages = "043",
    year = "2020"
}

@article{Bilicki:2012,
    author = "Bilicki, Maciej and Seikel, Marina",
    title = "{We do not live in the R\_h = c t universe}",
    eprint = "1206.5130",
    archivePrefix = "arXiv",
    primaryClass = "astro-ph.CO",
    doi = "10.1111/j.1365-2966.2012.21575.x",
    journal = "Mon. Not. Roy. Astron. Soc.",
    volume = "425",
    pages = "1664--1668",
    year = "2012"
}

@article{NanLin:2019,
    author = "Lin, Hai-Nan and Li, Xin and Tang, Li",
    title = "{Non-parametric reconstruction of dark energy and cosmic expansion from the Pantheon compilation of type Ia supernovae}",
    eprint = "1905.11593",
    archivePrefix = "arXiv",
    primaryClass = "gr-qc",
    doi = "10.1088/1674-1137/43/7/075101",
    journal = "Chin. Phys. C",
    volume = "43",
    number = "7",
    pages = "075101",
    year = "2019"
}

@article{Zhang16,
    author = "Zhang, Ming-Jian and Xia, Jun-Qing",
    title = "{Test of the cosmic evolution using Gaussian processes}",
    eprint = "1606.04398",
    archivePrefix = "arXiv",
    primaryClass = "astro-ph.CO",
    doi = "10.1088/1475-7516/2016/12/005",
    journal = "JCAP",
    volume = "12",
    pages = "005",
    year = "2016"
}

@article{Haridasu18,
    author = "Haridasu, Balakrishna S. and Lukovi\'c, Vladimir V. and Moresco, Michele and Vittorio, Nicola",
    title = "{An improved model-independent assessment of the late-time cosmic expansion}",
    eprint = "1805.03595",
    archivePrefix = "arXiv",
    primaryClass = "astro-ph.CO",
    doi = "10.1088/1475-7516/2018/10/015",
    journal = "JCAP",
    volume = "10",
    pages = "015",
    year = "2018"
}

@article{Mukherjee20,
    author = "Mukherjee, Purba and Banerjee, Narayan",
    title = "{Revisiting a non-parametric reconstruction of the deceleration parameter from combined background and the growth rate data}",
    eprint = "2007.15941",
    archivePrefix = "arXiv",
    primaryClass = "astro-ph.CO",
    doi = "10.1016/j.dark.2022.100998",
    journal = "Phys. Dark Univ.",
    volume = "36",
    pages = "100998",
    year = "2022"
}

@article{Mukherjee21,
	doi = {10.1140/epjc/s10052-021-08830-5},
  
	url = {https://doi.org/10.1140%2Fepjc%2Fs10052-021-08830-5},
  
	year = 2021,
	month = {jan},
  
	publisher = {Springer Science and Business Media {LLC}
},
  
	volume = {81},
  
	number = {1},
  
	author = {Purba Mukherjee and Narayan Banerjee},
  
	title = {Non-parametric reconstruction of the cosmological jerk parameter},
  
	journal = {The European Physical Journal C}
}

@article{SeikelEtAl12b,         
    author = "Seikel, Marina and Yahya, Sahba and Maartens, Roy and Clarkson, Chris",
    title = "{Using H(z) data as a probe of the concordance model}",
    eprint = "1205.3431",
    archivePrefix = "arXiv",
    primaryClass = "astro-ph.CO",
    doi = "10.1103/PhysRevD.86.083001",
    journal = "Phys. Rev. D",
    volume = "86",
    pages = "083001",
    year = "2012"
}

@article{YuEtAl18,    
    author = "Yu, Hai and Ratra, Bharat and Wang, Fa-Yin",
    title = "{Hubble Parameter and Baryon Acoustic Oscillation Measurement Constraints on the Hubble Constant, the Deviation from the Spatially Flat \ensuremath{\Lambda}CDM Model, the Deceleration\textendash{}Acceleration Transition Redshift, and Spatial Curvature}",
    eprint = "1711.03437",
    archivePrefix = "arXiv",
    primaryClass = "astro-ph.CO",
    doi = "10.3847/1538-4357/aab0a2",
    journal = "Astrophys. J.",
    volume = "856",
    number = "1",
    pages = "3",
    year = "2018"
}

@article{JesusEtAl20GP,
    author = "Jesus, J. F. and Valentim, R. and Escobal, A. A. and Pereira, S. H.",
    title = "{Gaussian Process Estimation of Transition Redshift}",
    eprint = "1909.00090",
    archivePrefix = "arXiv",
    primaryClass = "astro-ph.CO",
    doi = "10.1088/1475-7516/2020/04/053",
    journal = "JCAP",
    volume = "04",
    pages = "053",
    year = "2020"
}

@article{BenndorfEtAl22,
    author = "Benndorf, D. and Jesus, J. F. and Pereira, S. H.",
    title = "{Determination of the kinematic parameters from SNe Ia and cosmic chronometers}",
    eprint = "2109.09542",
    archivePrefix = "arXiv",
    primaryClass = "astro-ph.CO",
    doi = "10.1140/epjc/s10052-022-10378-x",
    journal = "Eur. Phys. J. C",
    volume = "82",
    number = "5",
    pages = "457",
    year = "2022"
}

@article{MorescoEtAl22, 
    author = "Moresco, Michele and others",
    title = "{Unveiling the Universe with emerging cosmological probes}",
    eprint = "2201.07241",
    archivePrefix = "arXiv",
    primaryClass = "astro-ph.CO",
    doi = "10.1007/s41114-022-00040-z",
    journal = "Living Rev. Rel.",
    volume = "25",
    number = "1",
    pages = "6",
    year = "2022"
}

@article{JesusEtAl22GP,
    author = "Jesus, J. F. and Valentim, R. and Escobal, A. A. and Pereira, S. H. and Benndorf, D.",
    title = "{Gaussian processes reconstruction of the dark energy potential}",
    eprint = "2112.09722",
    archivePrefix = "arXiv",
    primaryClass = "astro-ph.CO",
    doi = "10.1088/1475-7516/2022/11/037",
    journal = "JCAP",
    volume = "11",
    pages = "037",
    year = "2022"
}

@article{Riess:2016jrr,
    author = "Riess, Adam G. and others",
    title = "{A 2.4\% Determination of the Local Value of the Hubble Constant}",
    eprint = "1604.01424",
    archivePrefix = "arXiv",
    primaryClass = "astro-ph.CO",
    doi = "10.3847/0004-637X/826/1/56",
    journal = "Astrophys. J.",
    volume = "826",
    number = "1",
    pages = "56",
    year = "2016"
}








\bsp	
\label{lastpage}
\end{document}